\documentclass[amsmath,nofootinbib,nobibnotes,amssymb,aps,prl,superscriptaddress,twocolumn,floatfix, longbibliography]{revtex4-2}

\newcommand{\para}{}

\usepackage{float}
\usepackage{comment}
\usepackage{booktabs}
\usepackage[percent]{overpic}
\usepackage{microtype}
\usepackage{array}
\usepackage{amsmath}
\usepackage{amsthm}
\usepackage{amssymb}
\usepackage{physics}
\usepackage{graphicx}

\theoremstyle{remark}

\theoremstyle{definition}

\usepackage{bbold}
\numberwithin{thm}{section}
\usepackage{hyperref}
\usepackage[usenames,dvipsnames]{xcolor}
\usepackage{tikz}
\usetikzlibrary{arrows.meta, automata, positioning, quotes, shapes}
\usepackage{soul}

\newcommand{\xCornell}{Department of Physics, Cornell University, Ithaca, NY, USA}
\newcommand{\xEwha}{Department of Physics, Ewha Womans University, Seoul, South Korea}
\newcommand{\xHarvard}{Department of Physics, Harvard University, Cambridge, MA, USA}
\newcommand{\xRadcliffe}{Radcliffe Institute for Advanced Studies, Cambridge, MA, USA}
\newcommand{\xGoogle}{Google Research, Mountain View, CA, USA}

\begin{document}

\title{Machine learning reveals features of spinon Fermi surface}
\author{Kevin Zhang}
\email[Corresponding author: ]{kz345@cornell.edu}
\affiliation{\xCornell}
\author{Shi Feng}
\affiliation{Department of Physics, The Ohio State University, Columbus, OH, USA}
 \author{Yuri D. Lensky}
\affiliation{\xCornell}
\affiliation{\xGoogle}
\author{Nandini Trivedi}
\affiliation{Department of Physics, The Ohio State University, Columbus, OH, USA}
 \author{Eun-Ah Kim}
\affiliation{\xCornell}
\affiliation{\xRadcliffe}
\affiliation{\xHarvard}
\affiliation{\xEwha}

\def\bibsection{\section*{\refname}} 

\begin{abstract}
{\bf Abstract.} \\
With rapid progress in simulation of strongly interacting quantum Hamiltonians, the challenge in characterizing unknown phases becomes a bottleneck for scientific progress.
We demonstrate that a Quantum-Classical hybrid approach (QuCl) of mining sampled projective snapshots with interpretable classical machine learning can unveil signatures of seemingly featureless quantum states.
The Kitaev-Heisenberg model on a honeycomb lattice under external magnetic field presents an ideal system to test QuCl, where simulations have found an intermediate gapless phase (IGP) sandwiched between known phases, launching a debate over its elusive nature.
We use the correlator convolutional neural network, trained on labeled projective snapshots, in conjunction with regularization path analysis to identify signatures of phases.
We show that QuCl reproduces known features of established phases.
Significantly, we also identify a signature of the IGP in the spin channel perpendicular to the field direction, which we interpret as a signature of Friedel oscillations of gapless spinons forming a Fermi surface.
Our predictions can guide future experimental searches for spin liquids.

\end{abstract}
\date{\today}
\maketitle

\para {\bf Introduction.} \\
As our ability to simulate quantum systems increases, there is a corresponding need for determining how to characterize unknown phases realized in simulators.
Going from measurements to the nature of the underlying state is a challenging inverse problem.
Full quantum state tomography \cite{James2001Phys.Rev.A} of the density matrix is impractical.
Although the classical shadow \cite{Huang2020Nat.Phys.} scales better than full tomography, the approach does not prescribe to researchers the proper observables to evaluate.
Viewing the inverse problem as a data problem invites adopting machine learning methods: a quantum-classical hybrid approach. 
Machine learning has been widely applied for characterizing quantum states~\cite{Carrasquilla2020Adv.Phys.X}.
Such methods have been most fruitful with symmetry-broken states, with a diverse set of approaches increasingly bringing more interpretability and reducing bias~\cite{Miles2021NatCommun,Arnold2022Phys.Rev.X,Miles2023Phys.Rev.Res.}. 
The characteristic features of ordered phases are ultimately local and classical, hence ML models tuned for image processing have readily learned such features. 
By contrast, past learning of quantum states defined without order parameters has relied on theoretically guided feature preparation~\cite{Zhang2017Phys.Rev.B,Huang2022Science}.
However, such reliance on prior knowledge blocks the researchers' access to new insights into unknown states: the ultimate goal of simulating quantum states.

\para To push the limits of the nascent quantum-classical hybrid approach, we need a setting 
known to host a non-trivial quantum phase of unknown nature.
Recent investigations into extended Kitaev models \cite{Hermanns2018Annu.Rev.Condens.MatterPhys.,Knolle2019Annu.Rev.Condens.MatterPhys.,Gohlke2017Phys.Rev.Lett.,Trebst2022PhysicsReports} have led to the observation of a mysterious intermediate gapless phase (IGP) sandwiched between the Kitaev spin liquid and the trivial polarized state under a non-perturbative [111] magnetic field \cite{Zhu2018Phys.Rev.B,Gohlke2018Phys.Rev.B,Jiang2019Phys.Rev.B,Ronquillo2019Phys.Rev.B}, whose identification presents an interesting and important puzzle away from the perturbative limit.
However, the nature of this field-induced IGP has raised debate in the community.

\para Several theories have shown evidence that supports a gapless quantum spin liquid phase with an emergent $U(1)$ spinon Fermi surface ~\cite{Jiang2018,Hickey2019NatCommun,Patel2019Proc.Natl.Acad.Sci.,Jiang2019Phys.Rev.B}, while there are also mean field theories indicating that the low energy effective theory of the intermediate phase is gapped with a non-zero Chern number \cite{Jiang2020Phys.Rev.Lett.,Zhang2022NatCommun}.  
This tension between theories arises due to the challenge in determining the nature of the IGP that forms in a non-perturbative region under a magnetic field.
Unlike the gapped topological phase adiabatically connected to the exactly solvable limit with known loop operators \cite{Zhang2017Phys.Rev.B,Feng2023a,Liu2021Phys.Rev.Res.}, 
the absence of measurable positive features for the possible candidate IGP states \cite{Patel2019Proc.Natl.Acad.Sci., Hickey2019NatCommun} also makes this problem a worthy challenge for machine learning.

\begin{figure*}[t]
    \includegraphics[width=0.96\textwidth]{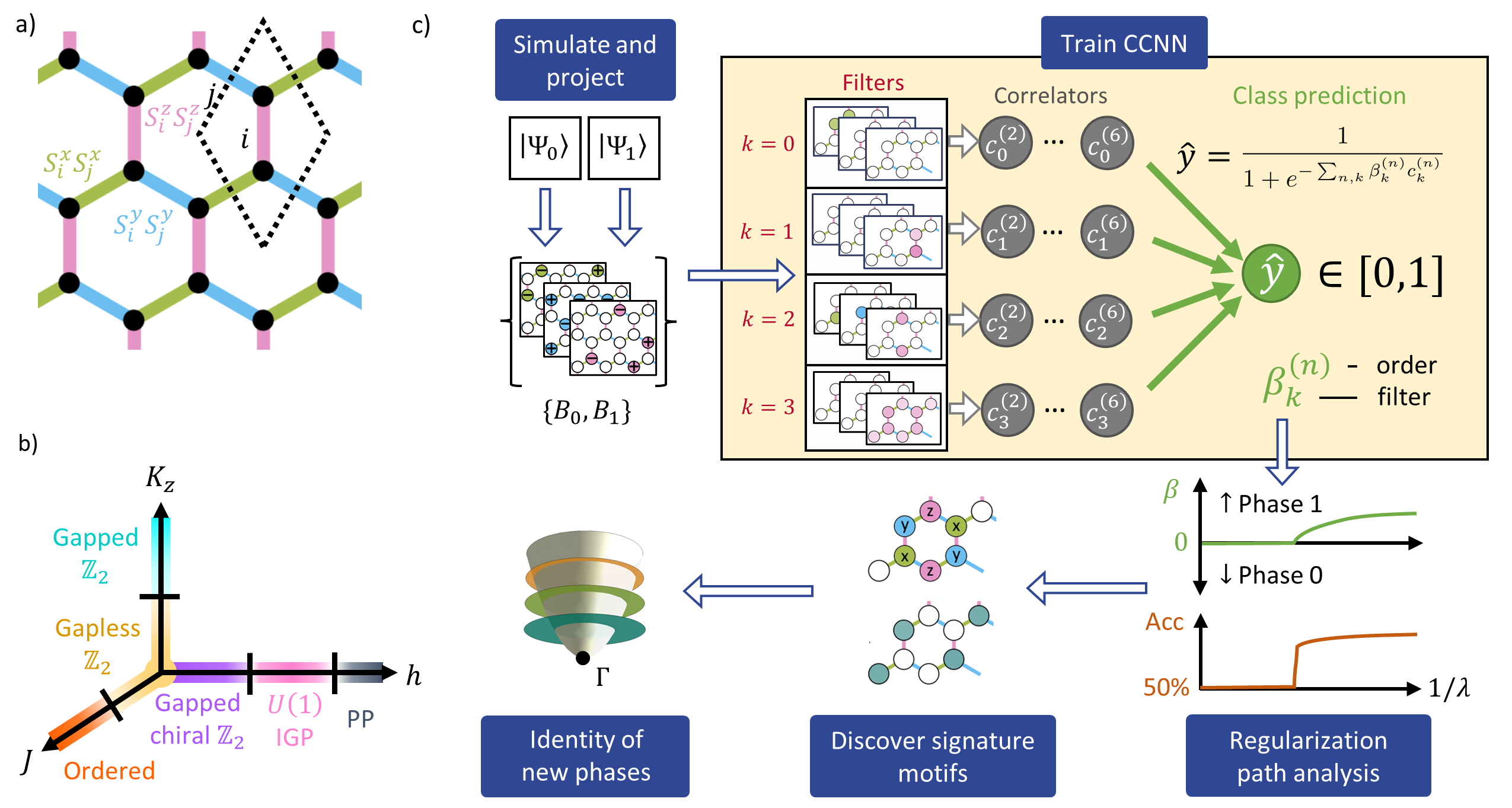}
    \caption{{\bf The Kitaev-Heisenberg model and schematic of the Quantum-Classical (QuCl) approach.}
    a) Honeycomb lattice of Kitaev model with bond-dependent interactions indicated by the three different colored bonds.
    b) Phase diagram of Kitaev-Heisenberg model in an external magnetic field (Eq~\eqref{eq:model})  long three axes of $K_z$, $h$, and $J$.
    c) A schematic description of QuCl: (i) From a pair of variational wavefunctions$|\Psi_0\rangle$ and $|\Psi_1\rangle$, labeled projective measurements (``snapshots") $B_0$ and $B_1$ are generated. (ii) The collection of labeled snapshots are used to train the correlator convolutional neural network (CCNN). (iii) The CCNN is configured with four filters $(k=0,\cdots,3)$ each with three channels, for the binary classification problem minimizing the distance between the prediction $\hat{y}$ and the label.
    (iv) Once the training is completed, we fix the filters and use regularization path analysis to reveal signature motifs of the two phases, 0 and 1, under consideration. 
    The correlator weight $\beta$ onsetting upon reduction of the regularization strength $\lambda$ to a negative (positive) value signals a feature of the phase 0 (1).}
    \label{fig:overview}
\end{figure*}

\para We present a quantum-classical hybrid approach, QuCl, to reveal characteristic motifs associated with states without known signature features.
We treat variational wavefunctions obtained from density matrix renormalization group (DMRG) \cite{White1992Phys.Rev.Lett.a,White1993Phys.Rev.B,Fishman2022SciPostPhys.Codebasesa} as output of a quantum simulator. Namely, we sample snapshots from the ground state and train an interpretable neural network architecture, i.e. the correlator convolutional neural network (CCNN)~\cite{Miles2021NatCommun} (\autoref{fig:overview}(c)).
Based on the trained network, we use regularization path analysis~\cite{Efron2004Ann.Stat.} to determine the distinct correlation functions learned by the CCNN as characteristic features of the state captured by snapshots.
We benchmark the performance of this hybrid approach on the known phases and confirm that the CCNN learned features are consistent with the known characteristic features. Importantly, we reveal the signature feature of the IGP to imply the existence of a spinon Fermi surface, as proposed in \cite{Patel2019Proc.Natl.Acad.Sci.,Hickey2019NatCommun}. 

\begin{figure}[t]
    \includegraphics[width=0.48\textwidth]{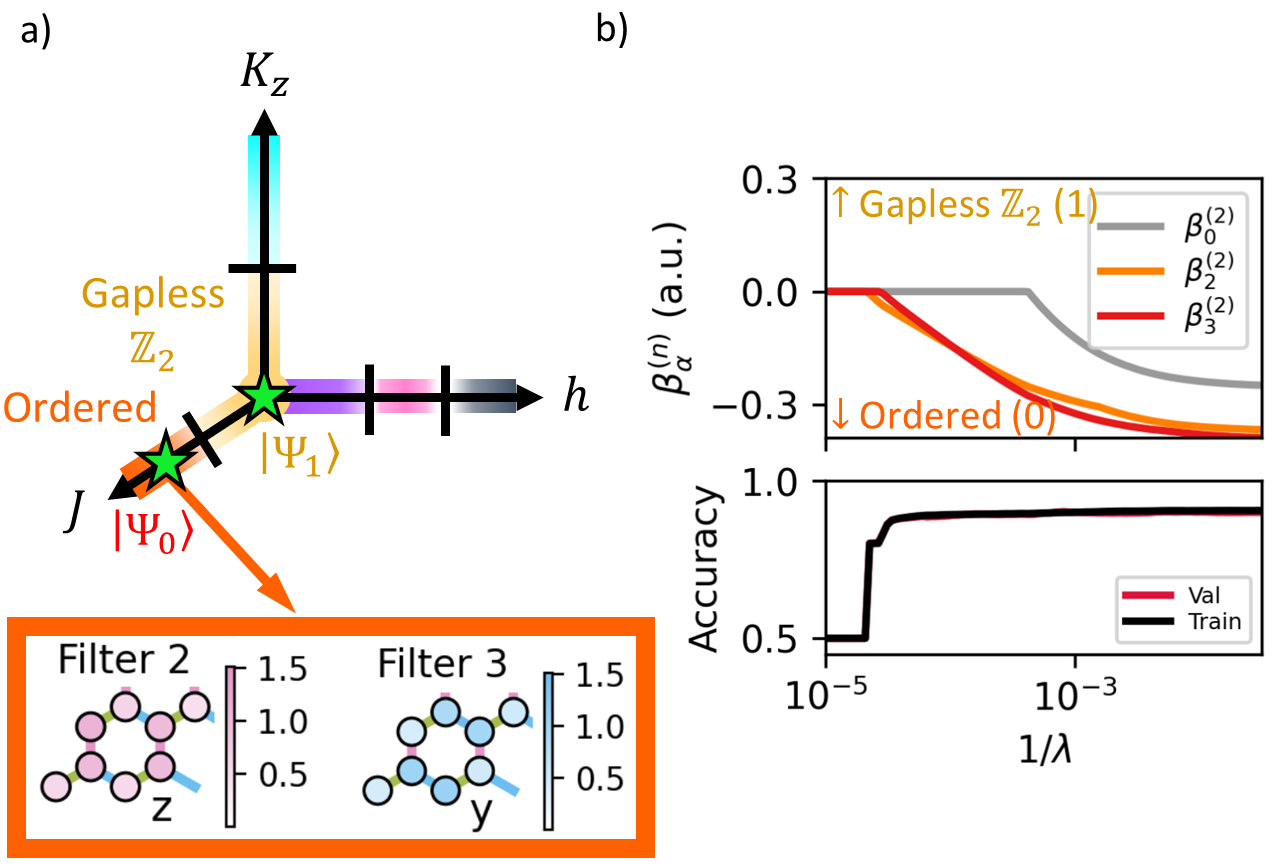}
    \caption{
    {\bf Gapless $\mathbb{Z}_2$ v.s. Heisenberg ordered phase.}
    a) Ground state wavefunctions from the gapless $\mathbb{Z}_2$ phase and ferromagnetically ordered phase are obtained at the points on the $J$-axis marked by stars. 
    The highlighted box shows the two most informative filters for the classification task. The pink and blue dots in filters correspond to projections in $z$ and $y$ basis. 
    b) The regularization path analysis results pointing to the filters in panel a as signature motifs of the ordered phase. 
    }
    \label{fig:gaplessheis}
\end{figure}

\para {\bf Results and Discussion.} \\
\textit{Model.}
The Kitaev-Heisenberg model under an external field is defined by
\begin{equation} \label{eq:model}
    H = \sum_{\gamma = x,y,z} \sum_{\langle i j \rangle_{\gamma}}  K_\gamma S^{\gamma}_i \ S^{\gamma}_j - J \sum_{\langle i j \rangle} \mathbf{S}_i \cdot \mathbf{S}_j - \sum_{i} \mathbf{h}\cdot\mathbf{S}_i 
\end{equation}
where $\gamma = x,y,z$ enumerates the three colorings of bonds on the honeycomb lattice (\autoref{fig:overview}(a)), and $S^\gamma$ is the $\gamma$ projection of spin-$1/2$ degrees of freedom on each site. We also add a uniform Zeeman field $\mathbf{h}$ along the $[111]$ direction, i.e., the out-of-plane $\hat{e}_3$ direction in the lab frame (see also Supplementary Note III), as well as a ferromagnetic Heisenberg term of strength $J$. Here we investigate the antiferromagnetic Kitaev intereaction ($K>0$), for which a field in the $[111]$ direction gives rise to an intermediate phase over a significant field regime. For the ferromagnetic Kitaev interaction, on the other hand, the intermediate phase is either absent or exists over a very small field regime~\cite{Hickey2019NatCommun}. 
Starting from the exactly solvable point at $K_x = K_y = K_z = 1$, $h=J=0$, which is a nodal $\mathbb{Z}_2$ spin liquid \cite{Kitaev2006AnnalsofPhysics},
we consider three axes of the phase diagram that are controlled by the parameters $h$, $J$, and $K_z$ (\autoref{fig:overview}(b)).

\para For the $J$ axis of the phase diagram in \autoref{fig:overview}(b), the system undergoes a sequence of transitions through magnetically ordered states~\cite{Gohlke2017Phys.Rev.Lett.}.
For small values of $J$ the system 
preserves time-reversal symmetry and the system remains a gapless $\mathbb{Z}_2$ spin liquid.
As $J$ is increased, the system acquires a zigzag magnetic order (also experimentally observed in $\alpha$-RuCl$_3$ \cite{Jackeli2009Phys.Rev.Lett.,Gohlke2017Phys.Rev.Lett.,Kasahara2018Nature}).
At even larger values of $J$, the system eventually becomes a Heisenberg ferromagnet.
On the other hand, a small magnetic field $h\parallel[111]$ breaks the time reversal symmetry of the Kitaev model and opens a gap in the spectrum of free majorana fermions, resulting in a CSL~\cite{Kitaev2006AnnalsofPhysics}.
However, upon leaving the perturbative regime, numerical evidence through DMRG \cite{Patel2019Proc.Natl.Acad.Sci.} and exact diagonalization \cite{Ronquillo2019Phys.Rev.B,Hickey2019NatCommun} have shown that the system goes through an IGP before entering a partially polarized (PP) magnetic phase.
Although the precise nature of the IGP is unknown, a $U(1)$ spinon Fermi surface has been proposed recently \cite{Hickey2019NatCommun,Patel2019Proc.Natl.Acad.Sci.}, which, as we are to show, is in agreement with our CCNN results.
Finally, the Kitaev model has an exact solution along the $K_z$ axis when $J=h=0$~\cite{Kitaev2006AnnalsofPhysics}, where the system undergoes a transition to a gapped $\mathbb{Z}_2$ spin liquid upon increasing $K_z$. We use this axis for benchmarking the QuCl outcome to known exact results. 

\para In order to generate a single snapshot from a wavefunction, we perform the following procedure sequentially on each site $i$ of the lattice:
\begin{enumerate}
    \item Find the reduced density matrix for site $i$, and exactly evaluate the expectation value of the spin operator projected along the chosen axis $\alpha_i$.
    \item Choose eigenvalue $+$ or $-$ with probabilities $P_+ = \frac{1+\langle\sigma^\alpha_i\rangle}{2}$, $P_- = 1-P_+$; record the eigenvalue and axis of projection.
    \item Collapse the wavefunction onto the associated eigenstate of site $i$ using the projector $\ket{\pm \alpha}_i\bra{\pm \alpha}_i$.
    \item Repeat 1-3 until every site is addressed. 
    \item Organize the snapshots into channels, one for each unique axis $\alpha_i$; see \autoref{fig:overview}(c).
\end{enumerate}
The choice of axis $\alpha_i$ is random for the $J$ and $K_z$ axes but tailored to the target phases for the $h$ axis. 
The wavefunctions at phase space points of interest are obtained using DMRG on finite size systems composed of $6\times 5$-unit-cell (60 sites). In Supplementary Note III we also show results from an extended $12\times 3$-unit-cell (120 sites) cylinder geometry.
In both cases we used a maximum of 1200 states, giving converged results with a truncation error $\sim 10^{-7}$ or less in all phases. 
Within a phase, we generated 10,000 snapshots for each wavefunction in question.
Each resulting snapshot forms a three-dimensional array of bit-strings, with two spatial dimensions and a ``channel" dimension (see \autoref{fig:overview}(c)).
Such a collection of snapshots is a classical shadow of the quantum state~\cite{Huang2020Nat.Phys.}.
Since our goal is to characterize a quantum state without prior knowledge of the best operator to measure, we treat the snapshot collection as data rather than using them to estimate an operator expectation value as in Refs.~\cite{Huang2020Nat.Phys.,Ferris2012Phys.Rev.Ba}.

\begin{figure*}[t]
    \includegraphics[width=0.96\textwidth]{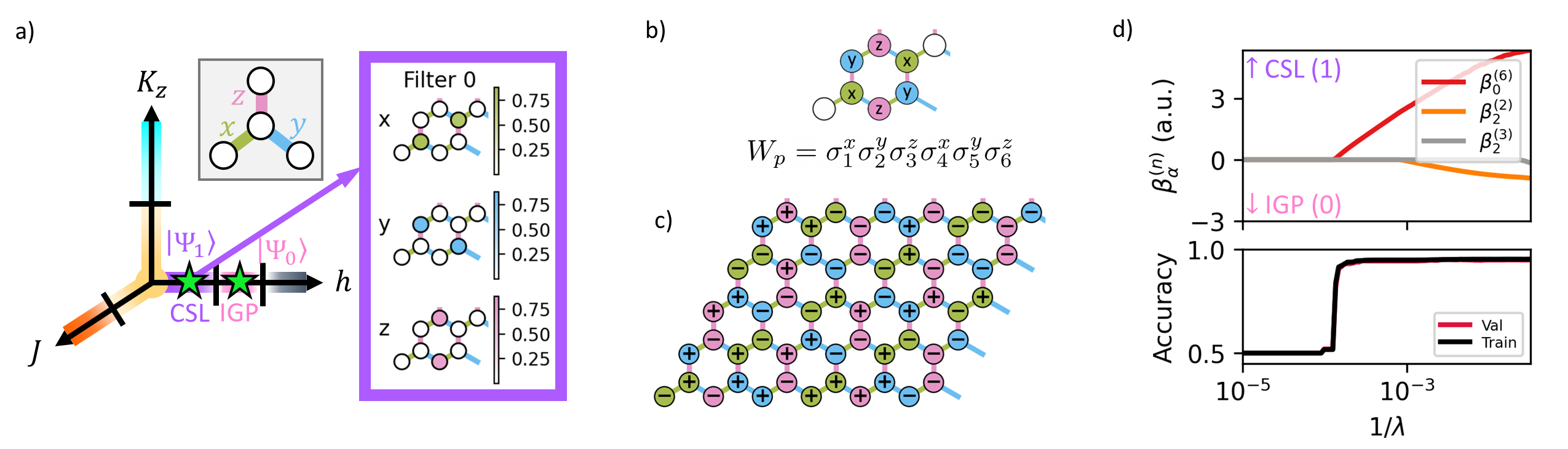}
    \caption{{\bf The chiral spin liquid v.s. the intermediate gapless phase, benchmarking the indicator of the chiral spin liquid phase.} 
    a) Ground state wavefunctions from the chiral spin liquid (CSL) phase and the intermediate gapless phase (IGP) are obtained at the points on the $h$-axis marked by stars. 
    The highlighted box shows the most informative filter that signifies the CSL phase. 
    Inset shows the bond anisotropy for the three colorings of bonds, e.g. $x$ refers to a $S^x_iS^x_j$ coupling.
    b) The plaquette operator $W_p$ is a operator defined on the six sites around a hexagonal plaquette.
    c) A sample snapshot from the CSL phase showing the measurement basis that makes the plaquette operator $W_p$ accessible.
    d) The regularization path analysis pointing to the six-point correlator of filter in panel a as indicator of the CSL.}
    \label{fig:csligp}
\end{figure*}

\para For each axis of the phase diagram, we set up a binary classification problem between a pair of phase space points, $|\Psi_0\rangle$ and $|\Psi_1\rangle$, each deep within a phase. 
The machine learning architecture of choice, CCNN, 
was introduced in Ref.~\cite{Miles2021NatCommun} as an adaptation of a convolutional neural network where a controlled polynomial non-linearity splits into different orders of correlators for the neural network to use~(\autoref{fig:overview}(c)).
Compared to the more standard CNN architecture, the CCNN has reduced expressibility due to using a low-order polynomial as the nonlinearity.
However, at the expense of this reduction, we gain access to interpreting the network's learning that can be analytically connected to the traditional notion of correlation functions.
Combined with regularization path analysis (RPA; see Methods and Supplementary Note I for details)~\cite{Tibshirani1996J.R.Stat.Soc.Ser.BMethodol.}, the CCNN can reveal spatial correlations or motifs that are characteristic of a given phase.

\para For a given channel $\alpha$ of filter $k$ to be learned, $f_{k,\alpha}$,
the CCNN samples correlators for each snapshot bit-string $B^\alpha(\mathbf{x})$ through an estimate for the $n$-th order spatially averaged correlator associated with filter $k$
\begin{equation}
    c_k^{(n)} = \sum_{\mathbf{x}} \sum_{(\mathbf{a_1}, \alpha_1)\neq ... \neq (\mathbf{a_n}, \alpha_n)}  \prod_{j=1}^{n} f_{k,\alpha_j}(\mathbf{a_j}) B^{\alpha_j}(\mathbf{x}+\mathbf{a_j}),
\end{equation}
where the inner sum is over all $n$ unique pairs of filter positions $\mathbf{a}$ and filter channels $\alpha$.
These correlator estimates are then coupled to coefficients $\beta_{k}^{(n)}$ of the linear layer (\autoref{fig:overview}(c); green arrows) according to $\hat{y} = \left[1 + \exp(-\sum_{n,k} \beta_k^{(n)} c_k^{(n)})\right]^{-1}$, where $0 \leq \hat{y} \leq 1$ is the CCNN output for the given input snapshot. We reserved 1,000 samples from each wavefunction as a validation set, and used the remaining 9,000 for training.
The orders of correlators were restricted to be between 2 and 6, inclusive.
We allowed the neural network to learn up to 4 different filters, corresponding to $0 \leq k \leq 3$.
The training optimizes the model parameters, namely the filters and the weights, by comparing the output $\hat{y}$ to the training label (see Methods).

\para Once the CCNN is successfully trained for a given phase, we uncover the characteristic motif that is most informative for the contrast using RPA~\cite{Tibshirani1996J.R.Stat.Soc.Ser.BMethodol.}. For this, we fix the filters and relearn the weights of each learned correlation $\beta_k^{(n)}$ with regularization that penalizes the magnitude of the $\beta_k^{(n)}$'s with strength $\lambda$ (see Methods).
The $\beta_k^{(n)}$ that turns on at the lowest value of $1/\lambda$ points to the specific filter $k$ and the correlation order $(n)$ of that filter which is most informative for the contrast task.
The sign of the onsetting $\beta_k^{(n)}$ reveals whether the associated correlation is a feature of phase 0 ($-$ sign) or of phase 1 ($+$ sign); see \autoref{fig:overview}(c).

\para \textit{Gapless $\mathbb{Z}_2$ versus Heisenberg phases.} As a benchmark, we first focus on the phases along the $J$ axis (\autoref{fig:gaplessheis}(a)).
At intermediate $J$, the system has zigzag order, while it is a Heisenberg ferromagnet for large $J$.
We trained the CCNN to distinguish wavefunctions from the two points marked by stars in \autoref{fig:gaplessheis}(a), at $J/K=0$ and $K/J=0$, corresponding to the Kitaev spin liquid and Heisenberg ferromagnetic states, respectively.
The snapshots were generated by choosing a random axis from $x$, $y$, or $z$ for each site.
The RPA shown in \autoref{fig:gaplessheis}(b) reveals that the most informative correlation functions are the two-point functions of filters 2 and 3 presented in \autoref{fig:gaplessheis}(a).
The negative sign of the onsetting $\beta$'s means these features are positive indicators of the ordered phase (see Supplementary Note I). Given that the correlation length vanishes at the exactly solvable point at the origin (phase 0), the network's choice to focus on features of phase 1 is sensible. Moreover, the learned motif of phase 1 is clearly consistent with a ferromagnetic correlation. Hence this benchmarking confirms that the CCNN's learning is consistent with our theoretical understanding when both phases 0 and 1 are known.

\para \textit{Chiral spin liquid.} Next, we contrast the CSL phase (phase 1) and the IGP (phase 0) along the $h$ axis (\autoref{fig:csligp}(a)). Neither of these phases is characterized by a local order parameter. However, the chiral phase is known to be a $\mathbb{Z}_2$ quantum spin liquid characterized by non-local 
Wilson loop expectation values \cite{Kitaev2006AnnalsofPhysics}. To confirm that such non-local information can be learned with our architecture, we first use snapshots with a fixed basis shown in \autoref{fig:csligp}(c) so that the architecture can access the necessary information.  
The RPA  with the positive onset of $\beta_0^{(6)}$ (\autoref{fig:csligp}(d)) implies that a sixth-order correlator of the filters shown in \autoref{fig:csligp}(a) is learned to be the key indicator of phase (1), the CSL phase. Remarkably, the relevant correlator $\langle \sigma_1^z \sigma_2^x \sigma_3^y \sigma_4^z \sigma_5^x \sigma_6^y \rangle$ is exactly the expectation value of the Wilson loop associated with the plaquette $p$ consisting of the six sites $\langle W_p \rangle$, shown in \autoref{fig:csligp}(b). Theoretically, $\langle W_p \rangle \approx 1$ implies the state is well-described by the $\mathbb{Z}_2$ gauge theory of the zero-field gapless phase \cite{Kitaev2006AnnalsofPhysics}. The fact that none other than $\langle W_p \rangle$ was learned to contrast the CSL phase from the intermediate gapless phase reveals that the latter is a distinct state. 
However, discovering the indicator of the intermediate phase requires a different approach, as we discuss below. 

\begin{figure*}
    \includegraphics[width=0.96\textwidth]{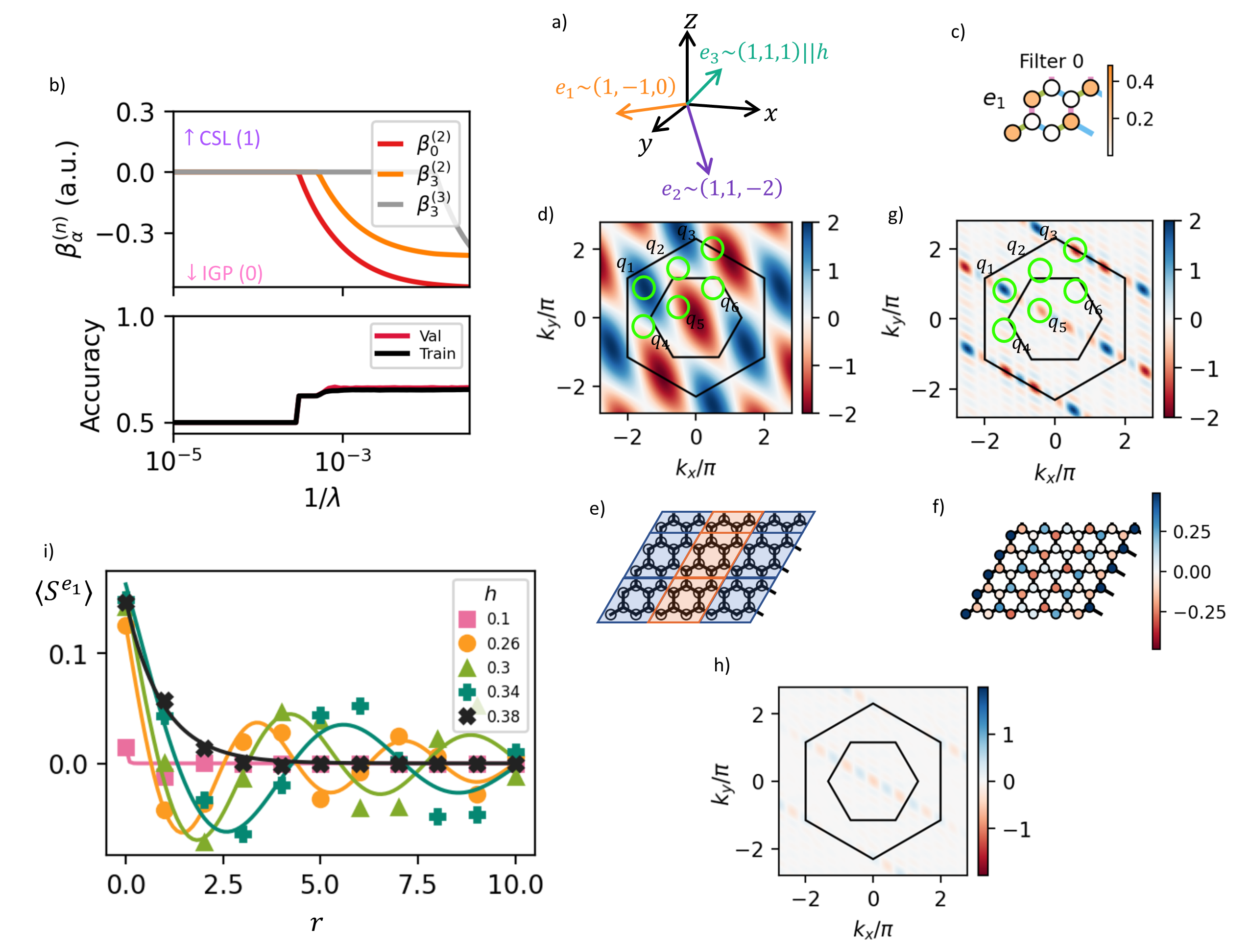}
    \caption{
    {\bf The chiral spin liquid v.s. the intermediate gapless phase, discovering the indicator of the intermediate gapless phase.} 
    a) The rotated basis vectors in relation to the cardinal axes; the external field is along $e_3$. The signature of the intermediate gapless phase (IGP) is targetted using $e_1$ basis snapshots from the same pair of wavefunctions as in \autoref{fig:csligp}(a).
    b) The regularization path analysis results pointing to the filter in panel c as an indicator of the IGP. 
    c) Most significant filter learned by the correlator convolutional neural network  to be associated with the IGP.
    d) Fourier transform of filter in panel c, with black lines indicating first and extended Brillouin zones. Green circles mark six Bragg peaks associated with the filter tiling pattern in panel e.
    e) Simplest possible tiling of filter shown in panel c, resulting in a superlattice of antiferromagnetic stripes. The Bragg peaks of this tiling pattern are marked by green circles in panels d and g.
    f) On-site magnetization $\expval*{S^{e_1}_i}$ of the wavefunction $|\Psi_0\rangle$ in the IGP.
    g) Real part of Fourier transform of panel f, again with Bragg peaks of antiferromagnetic tiling marked (imaginary part is negligible).
    h) Real part of Fourier transform of 
    $\expval*{S^{e_1}_i}$ of the wavefunction $|\Psi_1\rangle$ in the CSL phase showing no discernable features.
    i) The perpendicular magnetization $\expval*{S^{e_1}(r)}$ as a function of distance from boundary for various values of  field strength $h$, showing decreasing modulation period with increasing field strength. Solid lines show fitted curve based on \autoref{eq:osc}.}
    \label{fig:csligposcillation}
\end{figure*}

\para \textit{Intermediate Gapless Phase.}
We next discuss how we discover the physically meaningful features of the IGP (phase 0). Previous work has focused on mapping out the low energy excitations $S({\bf q},\omega\approx 0)$ in momentum space Also, in real space, the spin-spin correlations averaged over all directions shows power-law decay, indicating gapless spin excitations for intermediate fields. However, it has not been clear how to translate these correlations to positive signatures of a particular state that can be experimentally detected.
While the QuCl approach has the potential to reveal such signatures, we have to first overcome a ubiquitous challenge accompanying using ML for scientific discoveries: the need to guide the machine away from trivial features.
While the unbiased pursuit of representative feature in data is the benefit of using ML, a non-trivial cost is that the neural network's learning can be dominated by features that are trivial from the physicist's perspective.
The neural network's propensity to make decisions based on what appears most visible to the network means 
it is essential that we guide the CCNN away from the trivial yet dominant difference between phase 0 and phase 1: the field-driven magnetization along the $e_3$ axis (see Supplementary Note II). This basic requirement for extracting meaningful information using ML led us to supply CCNN with snapshots in the basis orthogonal to the field direction, such as $e_1$ basis (see \autoref{fig:csligposcillation}(a)).  
This decision to guide the CCNN away from trivial features led to a sought-after discovery.

\para The RPA shown in \autoref{fig:csligposcillation}(b) unambiguously points to two-point correlators of the filter shown in \autoref{fig:csligposcillation}(c) as a signature feature of the IGP.
As is clear from its Fourier transform shown in \autoref{fig:csligposcillation}(d), the filter implies the emergence of a length scale in the $e_1$ component of the magnetization.
Given that the $e_1$ direction is perpendicular to the direction of $h$-field, the repeating arrangement of the motif the filter is detecting must be anti-ferromagnetic.
One such ansatz we conjecture shown in \autoref{fig:csligposcillation}(e) will single out specific momenta points marked in \autoref{fig:csligposcillation}(d) from the Fourier intensity of the filter (see Supplementary Note IV for more details).
To confirm this conjecture, we explicitly measure the per-site $e_1$-magnetization, $\expval*{S^{e_1}(r)}$,  of the two states. The measurement outcome (\autoref{fig:csligposcillation}(f)) and its Fourier transform  (\autoref{fig:csligposcillation}(g))
confirms indeed the IGP state has a modulating $e_1$-magnetization that we inferred from the CCNN learned filter and the ansatz tiling the filter.
Furthermore, the contrast between Fourier transforms from the IGP \autoref{fig:csligposcillation}(g) and from the CSL \autoref{fig:csligposcillation}(h) establishes that the pattern and the associated length scale are unique features of the IGP. Remarkably, we find such modulation to be consistent with a conjecture~\cite{Jiang2019Phys.Rev.B,Ronquillo2019Phys.Rev.B,Jiang2018,Pradhan2020Phys.Rev.B,Patel2019Proc.Natl.Acad.Sci.,Hickey2019NatCommun,Feng2023a} that the IGP is a $U(1)$ spin liquid with a spinon Fermi surface. Note here that in a translationally invariant system the corresponding quantity is the two-point spin-spin correlation function $\expval*{S^{e_1}(0)S^{e_1}(r)}$ and its Fourier transform.  

\begin{figure*}[t]
    \includegraphics[width=0.96\textwidth]{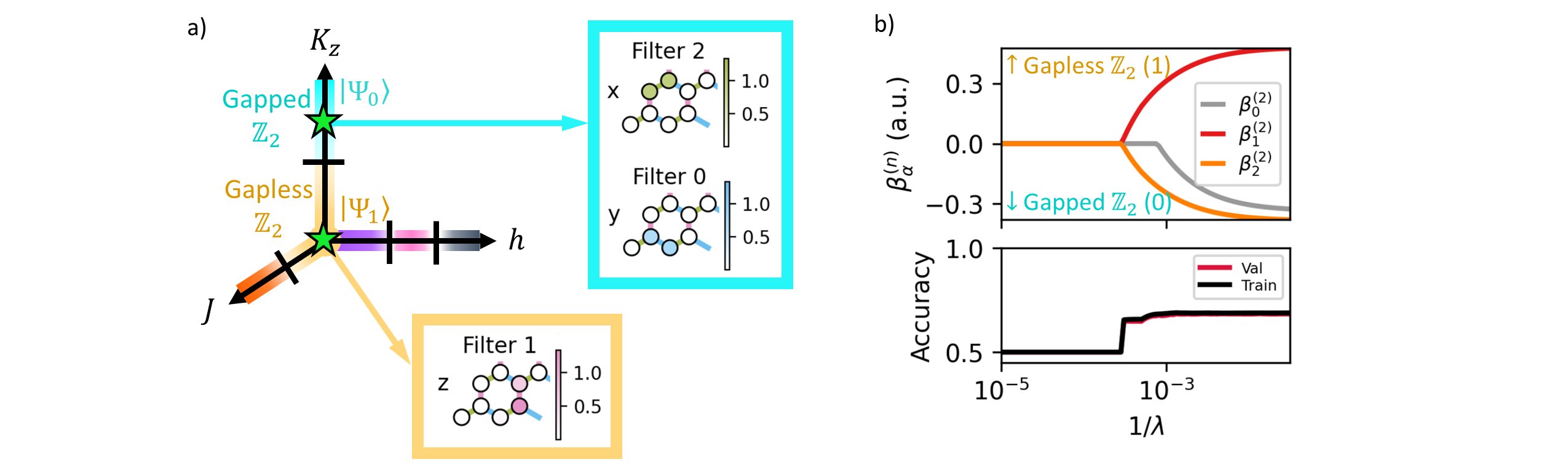}
    \caption{
    {\bf Gapless v.s. gapped $\mathbb{Z}_2$ spin liquid, benchmarking distinguishing two spin liquids.}
    a) Ground state wavefunctions from
the gapless $\mathbb{Z}_2$ phase and gapped $\mathbb{Z}_2$ phase are obtained at the points on the $K_z$-axis marked by stars. The highlighted boxes shows the most
informative filters that signifies the each phases.
    b) The regularization path analysis results associate two-point correlators of filters $0$ and $2$ to the gapped phase and that of filter $1$ to the gapless phase.}
    \label{fig:gaplessgapped}
\end{figure*}

\para As detailed in Supplementary Note III, $\expval*{S^{e_1}(r)}$ can be mapped to fermionic spinon density in the Kitaev model. If spinons are gapless and deconfined to form a spinon Fermi surface, the Friedel oscillation of the spinon density due to the open boundary \cite{White2002Phys.Rev.B,Mross2011Phys.Rev.Ba,He2018Phys.Rev.Lett.,Ruan2021Nat.Phys.b} will be reflected in the modulation of  $\expval*{S^{e_1}(r)}$:
\begin{equation} \label{eq:osc}
        \expval{S^{e_1}(r)}\sim \expval{n_1(r)} \sim \frac{k_F}{\pi}\left[ 1 - \frac{\sin(2k_F r+\theta)}{2k_F r+\theta} \right] + C
\end{equation}
where $C$ and $\theta$ are constants, and $r$ the distance measured from the boundary.
We confirm the spinon Friedel oscillation origin of the observed modulations by fitting the $\expval{S^{e_1}(r)}$ measured at different field strengths $h$ to Eq.~\eqref{eq:osc}. The resulting excellent fit in \autoref{fig:csligposcillation}(i)
shows that the modulation period increases with the increase in the perpendicular field $h$. This is consistent with a mean field picture in which the magnetic field plays the role of the chemical potential; the spinon bands successively get depleted upon increasing field until the system enters a trivial phase through a Liftshitz transition \cite{Feng2022Phys.Rev.B}. Evaluation of $S^{e_1}$ on a longer 20 unit cell system of a 3-leg ladder shows a modulation pattern that agrees with the 6-leg ladder results in \autoref{fig:csligposcillation}(i) (see Supplementary Note III).

\para \textit{Gapless $\mathbb{Z}_2$ vs gapped $\mathbb{Z}_2$.} Finally, we contrast the gapless and gapped $\mathbb{Z}_2$ phases along 
 the $K_z$ axis as a sanity check in distinguishing two spin liquid phases.
As one tunes $K_z$, the model Eq.~\eqref{eq:model} is known to go through a phase transition between a gapless $\mathbb{Z}_2$ and a gapped $\mathbb{Z}_2$ spin liquid phases \cite{Kitaev2006AnnalsofPhysics}.
However, since both phases have only short-range correlations in the ground state the distinction cannot be learned from the correlation lengths, unlike usual transitions between a gapless phase and a gapped phase. Hence it is a non-trivial benchmarking test for QuCl-based state characterization.
Contrasting the two points marked by stars in \autoref{fig:gaplessgapped}(a), again using the random basis snapshots, we find signature motifs consistent with exact solutions. Specifically, the RPA (\autoref{fig:gaplessgapped}(b)) shows that nearest-neighbor correlation functions of $x$ and $y$ axes are a feature of the gapped $\mathbb{Z}_2$ phase while the $z$ axis nearest-neighbor correlation function is the feature of the gapped $\mathbb{Z}_2$ phase.
These results are consistent with the exact solution of the zero-field Kitaev model \cite{Baskaran2007Phys.Rev.Lett.,Feng2022Phys.Rev.A}.

\para \textit{Conclusion.} The significance of our findings is threefold. Firstly, we gained insight into the intermediate field spin liquid phase in the Kitaev-Heisenberg model. 
Confronted by two complementary predictions: a gapless spin liquid based on exact diagonalization and DMRG versus a gapped spin liquid in the same region from mean field theory, an identification of a positive signature for either possibilities was critical.
The need for guiding the CCNN away from a trivially changing feature led to the discovery that it is critical to focus on snapshots taken along a direction $e_1$ perpendicular to the magnetic-field $e_3$ axis. 
Remarkably, the network then learned a geometric pattern characteristic of Friedel oscillations of spinons in the IGP.
This observation strongly supports earlier theoretical proposals of a spinon Fermi surface in the IGP, thus advancing our understanding of this phase.

\para Secondly, our discovery translates to a prediction for experiments by providing a 
direct evidence of spinon FS in the modulated magnetization and the spin-spin correlations perpendicular to the field direction along $e_1$. Such a feature in the computational data has been previously missed since the focus has been on isotropic spin correlation $\langle \vec{S}_i\cdot\vec{S}_j\rangle$ which is dominated by the $e_3$ component. Our results can guide future experimental searches for spin liquids with spinon Fermi surfaces. 

\para Finally, on a broader level, we have demonstrated that hidden features of a quantum many-body state can be discovered using QuCl: a data-centric approach to snapshots of the quantum states, employing an interpretable classical machine learning approach.
Conventionally, quantum states have been studied through explicit and costly evaluation of correlation functions.
However, when the descriptive correlation function is unknown in a new phase, the conventional approach gets lost in the overwhelming space of expensive calculations.
Although our method does not explicitly evaluate the correlation functions that it extracts, snapshots that can be readily treated with QuCl will enable computationally efficient identification of new phases associated with a quantum state, including topological states or states with hidden orders.
Finally, our method is also broadly applicable to searches for physical indicators of states prepared on quantum simulators which are naturally accessed through projective measurements.

\para {\bf Methods.} \\
In this section, we describe the architecture of the neural network and the training procedure.
The CCNN, as first proposed in Ref.~\cite{Miles2021NatCommun}, consists of two layers: the correlator convolutional layer and the fully connected linear layer.
We fed as input to the CCNN three-spin-channel (two-spin-channel for rotated basis measurements) snapshot data.
Since the CCNN was originally applied to square lattice data at its conception, we reinterpreted our hexagonal lattice geometry as a rectangular grid with a $1 \times 2$ unit cell forming its two-site basis.
We modified the convolutional layer to consist of 4 different learnable filters of dimension $2 \times 2$ unit cells, for a total receptive field of 8 sites each.
To accommodate the $1 \times 2$ unit cell, we also introduced a horizontal stride of 2 in the convolution operation between filters and snapshots.

\begin{figure*}
    \includegraphics[width=0.95\textwidth]{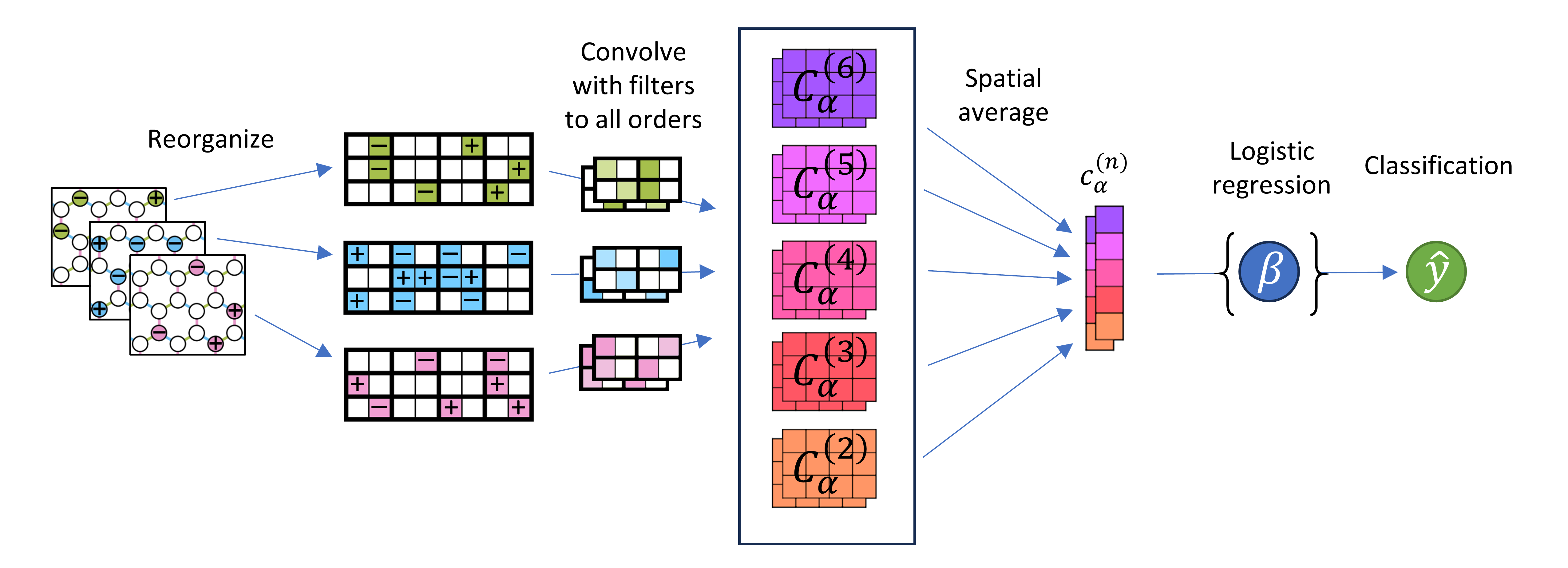}
    \caption{{\bf Visualization of correlator convolutional neural network architecture.}
    Input snapshots $B^\alpha(x)$ (with $\alpha$ enumerating the spin channels) are shaped into rectangular data arrays with a $1 \times 2$ unit cell.
    The inputs are convolved with filters to obtain orders $n = 2...6$ of correlations $C_k^{(n)}(x)$ (\autoref{eq:corr}), then spatially averaged to obtain correlator estimates $c_k^{(n)}$. Correlator estimates are then used in a logistic regression to predict the class of the input snapshot.}
    \label{fig:architecture}
\end{figure*}

The filter weights are learnable nonnegative numbers indicated by $f_{\alpha,k}(a)$, where $1 \leq \alpha \leq 4$ indexes the filter, $1 \leq k \leq 3$ indexes the channel of the weight, and $a$ is a spatial coordinate.
The weights are convolved with the input snapshots using the recursive algorithm described in Ref.~\cite{Miles2021NatCommun} to produce per-snapshot correlators as
\begin{equation}
    C_k^{(n)}(x) = \sum_{(\mathbf{a_1}, \alpha_1)\neq ... \neq (\mathbf{a_n}, \alpha_n)}  \prod_{j=1}^n f_{k,\alpha_j}(\mathbf{a_j}) B^{\alpha_j}(\mathbf{x}+\mathbf{a_j}),
    \label{eq:corr}
\end{equation}
where $C_k^{(n)}(x)$ is the position-dependent $n$-th order correlator of filter $k$, and $B^{\alpha_j}(\mathbf{x}+\mathbf{a_j})$ indicates the snapshot value at location $\mathbf{x}+\mathbf{a_j}$ in channel $\alpha_j$.
The correlator estimates are then defined as the spatially-averaged correlators, $c_k^{(n)} = \sum_x C_k^{(n)}(x)$, which are coupled to coefficients $\beta_{k}^{(n)}$ of the linear layer, and summed to produce the logistic regression classification output
\begin{equation}
    \hat{y} = \frac{1}{1 + \exp(-\sum_{n,k} \beta_k^{(n)} c_k^{(n)})}
\end{equation}
so that they are constrained to the range $0 \leq \hat{y} \leq 1$.
For a visual overview of the architecture, see \autoref{fig:architecture}.

During training, the weights of the network are updated with stochastic gradient descent to optimize the loss function
\begin{equation}
\begin{split}
L(y, \hat{y}) = -y \log \hat y - (1-y)\log(1-\hat y) \\
+ \gamma_1 \sum_{\alpha,k,a} f_{\alpha,k}(a) + \gamma_2 \sum_{\alpha,k,\bm a} f_{\alpha,k}(a)^2
\end{split}
\end{equation}
where $y \in \{0,1\}$ is the ground truth label of the snapshot, and $\gamma_1$ and $\gamma_2$ are L1 and L2 regularization strengths, respectively.
We took $\gamma_1=0.005$ and $\gamma_2=0.002$.
The training was performed for 20 epochs consisting of 9,000 snapshots each with a learning rate of 0.006, using Adam stochastic gradient descent.
For the regularization path analysis, the weights $f$ are kept fixed, and the model is retrained in the same way with loss function
\begin{equation}
L(y, \hat{y}) = -y \log \hat y - (1-y)\log(1-\hat y) + \gamma \sum_{k,n} \beta_k^{(n)},
\end{equation}
where $\gamma$ is the regularization strength to be swept over.

{\bf Data Availability.} Data is available upon request to the authors.

{\bf Code Availability.} Code is available upon request to the authors.

{\bf Acknowledgements.} We thank Leon Balents, Natasha Perkins, John Tranquada, and Simon Trebst for helpful discussions.
KZ acknowledges support by the NSF under EAGER OSP-136036 and the Natural Sciences and Engineering Research Council of Canada (NSERC) under PGS-D-557580-2021.
YL and EAK acknowledge support by the Gordon and Betty Moore Foundation’s EPiQS Initiative, Grant GBMF10436, and a New Frontier Grant from Cornell University’s
College of Arts and Sciences. EAK acknowledges support by the NSF under
OAC-2118310, EAGER OSP-136036, the Ewha Frontier 10-10 Research Grant, and the Simons
Fellowship in Theoretical Physics award 920665. 
SF acknowledges support from NSF Materials Research Science and Engineering Center (MRSEC) Grant No. DMR-2011876, and NT from NSF-DMR 2138905.

{\bf Author Contributions.} E-AK and NT conceived the idea and supervised the project. KZ, YL, and E-AK formed the machine learning and data processing strategies. SF performed the DMRG optimization and data processing. KZ performed the machine learning analysis. All authors contributed to interpreting the results and writing the manuscript.

{\bf Competing Interests.} The authors declare no competing interests.

\bibliography{biblio.bib}

\end{document}


\title{Supplementary Information for \\ ``Machine learning reveals features of spinon Fermi surface"}
\author{Kevin Zhang}
\affiliation{\xCornell}
\author{Shi Feng}
\affiliation{Department of Physics, The Ohio State University, Columbus, OH, USA}
 \author{Yuri D. Lensky}
\affiliation{\xCornell}
\affiliation{\xGoogle}
\author{Nandini Trivedi}
\affiliation{Department of Physics, The Ohio State University, Columbus, OH, USA}
 \author{Eun-Ah Kim}
\affiliation{\xCornell}
\affiliation{\xRadcliffe}
\affiliation{\xHarvard}
\affiliation{\xEwha}
\maketitle

\section{Supplementary Note I. Interpreting filters from regularization path analysis}

\begin{figure}
    \includegraphics[width=0.90\textwidth]{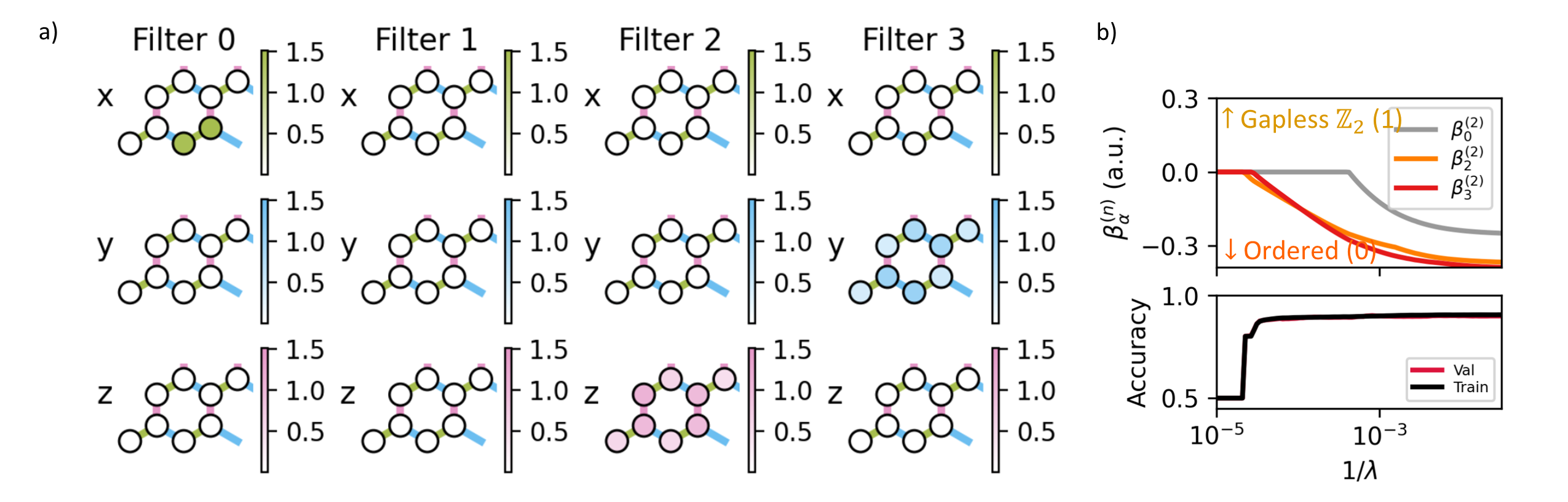}
    \caption{Interpretation of regularization path analysis in model distinguishing $Z_2$ gapless spin liquid from Heisenberg ferromagnet.
    a) All filters learned by neural network. Each column represents a separate filter, while the rows represent the different channels in the filters.
    b) Regularization path analysis showing relevant correlation functions used by neural network.}
    \label{fig:rpa}
\end{figure}

We now present an example of determining the most important filters from the learned weights and regularization path analysis.
For this example, we use the dataset and model described in the main text for distinguishing the $\mathbb{Z}_2$ gapless Kitaev spin liquid from the ferromagnetically ordered Heisenberg phase.
\autoref{fig:rpa}(a) shows all three channels of each of the four filters used in the convolutional layer of the neural network.
Notably, due to the L1 regularization imposed on the filter weights, most pixels are zero.
We now turn to the regularization path analysis to make sense of the filters.
In the large $\lambda$ limit, the accuracy is $50\%$, as the regularization penalty for having nonzero $\beta$'s outweighs the cross-entropy loss.
As $\lambda$ is decreased, the first two parameters that become nonzero are $\beta_2^{(2)}$ and $\beta_3^{(2)}$, which instructs us to look at Filter 2 and Filter 3.
The expansion of the second-order correlation functions is given by the sum of all pairs of two-point functions weighted by the product of the filter weights at both sites; in other words, for Filter 2,
\begin{equation}
    c_2^{(2)} = \sum_{\mathbf{a_1} \neq \mathbf{a_2}}  f_{k,e_z}(\mathbf{a_1}) B^{e_z}(\mathbf{x}+\mathbf{a_1}) f_{k,e_z}(\mathbf{a_2}) B^{e_z}(\mathbf{x}+\mathbf{a_2}),
\end{equation}
where the sum over channels is replaced by the single channel index $e_z$ by noting that the weights for the $x$ and $y$ channels of Filter 2 are zero.
Filter 3 can be treated in a corresponding way.
For Filter 2 and Filter 3 shown, the filter weights are approximately constant across all the pixels, which leads to
\begin{equation}
    c_2^{(2)} = \sum_{\mathbf{a_1} \neq \mathbf{a_2}} B^{e_z}(\mathbf{x}+\mathbf{a_1}) B^{e_z}(\mathbf{x}+\mathbf{a_2}),
\end{equation}
and is thus interpreted as the $q=0$ component of the spin-spin structure factor.
Finally, the sign of the $\beta_2^{(2)}$ and $\beta_3^{(2)}$ indicates the specific phase that the correlation functions are associated with: in this case, the signs being negative indicates that the filters point towards the ferromagnetic phase, which is consistent with the identification of the filters with the $q=0$ spin-spin correlation function.

\section{Supplementary Note II. Training in rotated basis}
 
\subsection{Parallel axis}

\begin{figure}
    \includegraphics[width=0.40\textwidth]{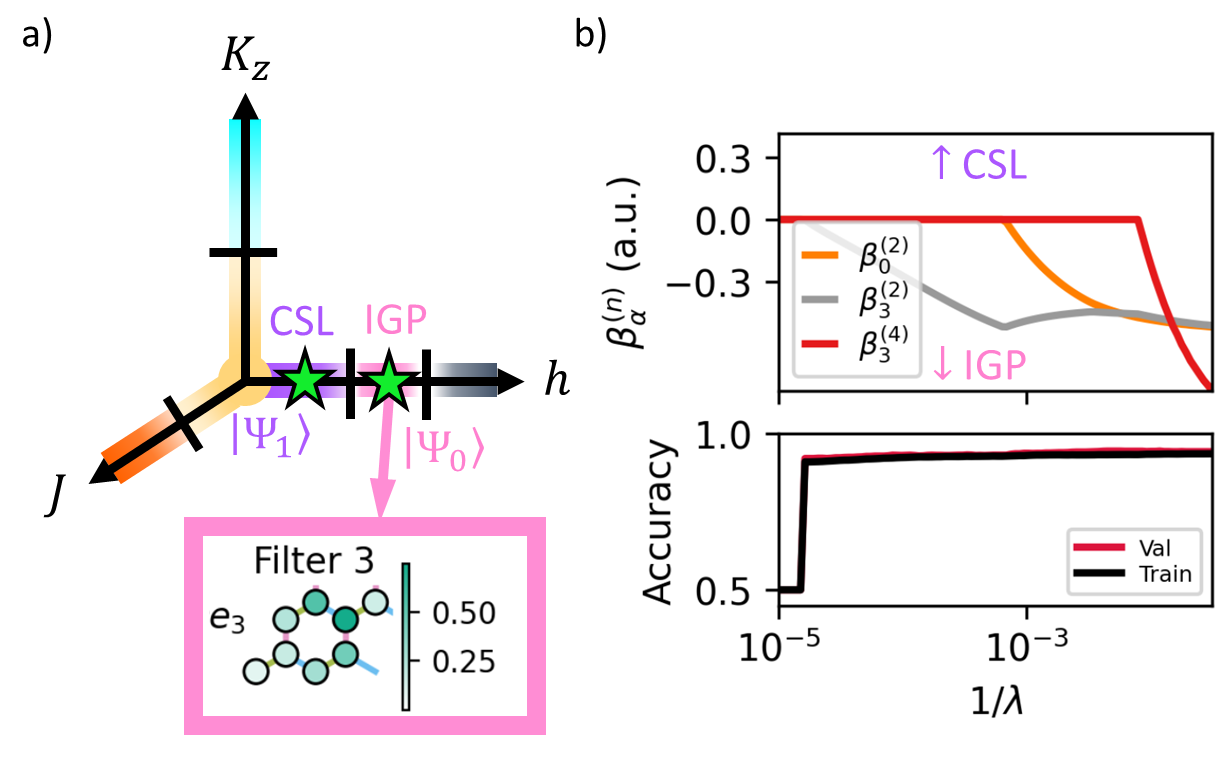}
    \caption{Details of the trained neural network that distinguishes the intermediate field gapless phase from the chiral spin liquid, using snapshots from the $e_3$ basis.
    a) Snapshots were obtained from wavefunctions in the CSL and intermediate field gapless phases.
    b) Regularization path analysis for the model in panel b.}
    \label{fig:e3}
\end{figure}

As a test case, we also trained the CCNN to distinguish the chiral spin liquid from the intermediate gapless phase using snapshots taken in the $e_3$ axis, which is parallel to the magnetic field.
Since the magnetization of the gapless phase is larger, the CCNN should be able to use this information to trivially distinguish the two phases.
Indeed, the two-point correlators of Filter 3 in \autoref{fig:e3}(a) were found to trivially characterize the gapless phase, as stated in the main text.
Since the CCNN estimates unconnected correlation functions, the larger $\langle S^{e_3} \rangle$ expectation value would cause all two-point functions of $S^{e_3}$ variables to be larger in the IGP, which is consistent with the learned correlation function.

\subsection{Perpendicular axis}

\begin{figure}
    \includegraphics[width=0.90\textwidth]{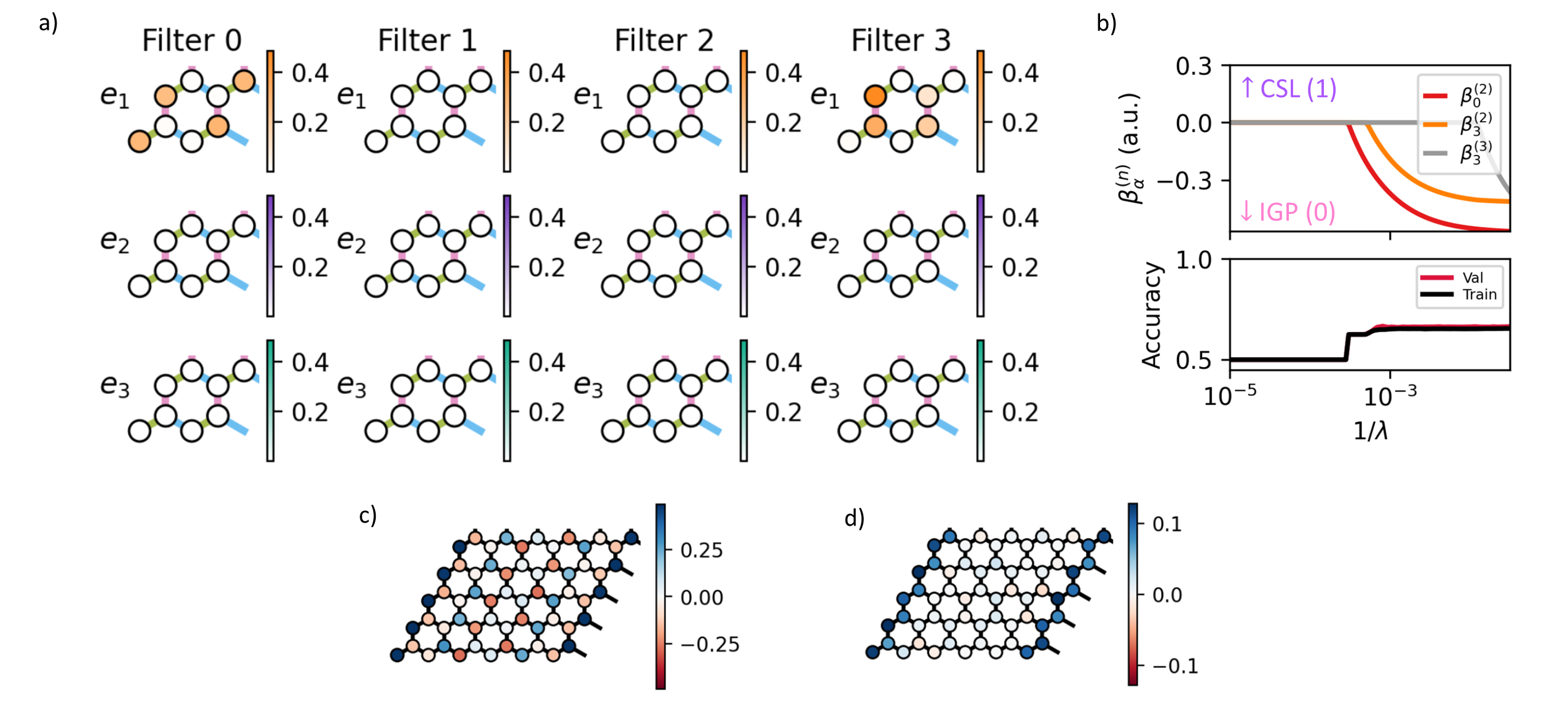}
    \caption{Interpretation of regularization path analysis in model distinguishing intermediate gapless phase from chiral spin liquid, using $e_1$ basis snapshots.
    a) All filters learned by neural network. Each column represents a separate filter, while the rows represent the different channels in the filters.
    b) Regularization path analysis showing relevant correlation functions used by neural network.
    c,d) Real-space magnetization in $e_1$ axis for gapless and CSL phases, respectively.}
    \label{fig:rpa2}
\end{figure}

Here, we present details of CCNN training using the $e_1$ basis snapshots \autoref{fig:rpa2}.
Since the snapshots were only taken in the $e_1$ basis, the $e_2$ and $e_3$ channels of the filters are left unused.
Regularization path analysis (\autoref{fig:rpa2}(b)) shows the most significant correlation function characterizing the IGP being the second-order correlators of Filter 0, as analyzed in the main text.

We note that the magnetization of the IGP (\autoref{fig:rpa2}(c)) is highly contrasted with the magnetization of the CSL phase, since the CSL magnetization is largely featureless except for edge effects.
On the other hand, the IGP magnetization is rich, and we provide an explanation of how Filter 0 predicts the observed pattern both in the main text, and also in SM VI.

\section{Supplementary Note III. Parton decomposition in rotated basis}
\begin{figure}
    \includegraphics[width=0.3\textwidth]{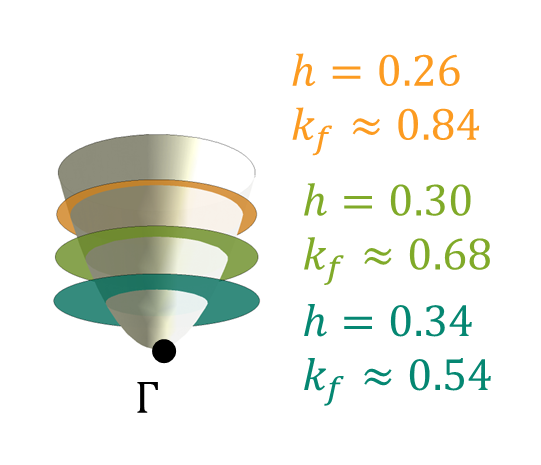}
    \caption{Cartoon picture of spinon band dispersion at various magnetic fields, with corresponding Fermi momentum at each point.}
    \label{fig:band}
\end{figure}

We first change the axis of spins to the $e_i$ basis which is relevant for the crystal symmetry of $\alpha$-RuCl$_3$ and other candidate Kitaev materials. The intrinsic spin axis $\hat{x}, \hat{y}, \hat{z}$ can be defined by the ligands that are responsible for generating the frustrated interaction due to spin-orbital coupling. However, it is more convenient to study under the lab coordinates consisting of an out-of-plane normal vector and two orthonormal in-plane vectors, namely $\hat{e}_1, \hat{e}_2$ and $\hat{e}_3$ unit vectors. The lab coordinate is related to the intrinsic coordiante via:
\begin{equation}
    \hat{e}_1 = \frac{1}{\sqrt{6}}(-\hat{x} - \hat{y} + 2\hat{z}),~ \hat{e}_2 = \frac{1}{\sqrt{2}}(\hat{x} - \hat{y}),~\hat{e}_3 = \frac{1}{\sqrt{3}}(\hat{x} + \hat{y} + \hat{z})
\end{equation}
which is readily written into the useful matrix form:
\begin{equation}
    \vec{\mathbf{R}} = 
    	\begin{pmatrix}
		-\frac{1}{\sqrt{6}} & -\frac{1}{\sqrt{6}} & \frac{2}{\sqrt{6}}\\
		\frac{1}{\sqrt{2}} & -\frac{1}{\sqrt{2}} & 0\\
		\frac{1}{\sqrt{3}} & \frac{1}{\sqrt{3}} & \frac{1}{\sqrt{3}}
	\end{pmatrix}
\end{equation}
which gives the new spin axis according to:
\begin{equation}
	\mathbf{S}' \equiv
	\begin{pmatrix}
		S^{e_1} \\ S^{e_2} \\ S^{e_3}
	\end{pmatrix}
	= \vec{\mathbf{R}}\mathbf{S} =
	\begin{pmatrix}
		-\frac{1}{\sqrt{6}} & -\frac{1}{\sqrt{6}} & \frac{2}{\sqrt{6}}\\
		\frac{1}{\sqrt{2}} & -\frac{1}{\sqrt{2}} & 0\\
		\frac{1}{\sqrt{3}} & \frac{1}{\sqrt{3}} & \frac{1}{\sqrt{3}}
	\end{pmatrix}
	\begin{pmatrix}
		S^x \\ S^y \\ S^z
	\end{pmatrix}
\end{equation}
where $\vec{\mathbf{R}}\in SO(3) \subset SU(2)$. 
Hence we can rewrite this in terms of $S^{e_1}, S^{e_2}, S^{e_3}$. After a spin rotation, they are related to the original operators by
\begin{align}
	S_i^{e_1} = \frac{1}{\sqrt{6}}(-S_i^x - S_i^y + 2S_i^z),~
	S_i^{e_2} = \frac{1}{\sqrt{2}}(S_i^x - S_i^y),~
	S_i^{e_3} = \frac{1}{\sqrt{3}}(S_i^x + S_i^y + S_i^z)
\end{align}
It is readily seen that these new operators satisfies SU(2) Lie algebra:
\begin{equation}
	[S_i^{e_1}, S_i^{e_2}] = iS^{e_3},~ [S_i^{e_2}, S_i^{e_3}] = i S_i^{e_1},~ [S_i^{e_3}, S_i^{e_1}] = iS_i^{e_2}
\end{equation}
Spin operators are decomposed into four majorana particles in Kitaev's exact solution \cite{Kitaev2006AnnalsofPhysics}. In our rotated basis we label these four majoranas as $c,b^{e_1}, b^{e_2}, b^{e_3}$. 
Now, let us write the majorana decomposition in the following way.
Define two fermion modes $\psi_1,~ \psi_2$ for each site, where
\begin{equation}
	c_i = (\psi_{i,1} + \psi_{i,1}^\dagger),~ b_i^{e_1} = i(\psi_{i,1}^\dagger - \psi_{i,1}),~ b_i^{e_2} = \psi_{i,2} + \psi_{i,2}^\dagger,~ b_i^{e_3} = i(\psi_{i,2}^\dagger - \psi_{i,2}),~  
\end{equation}
such that each spin is now
\begin{align}
	S_i^{e_1} &= ic_ib_i^{e_1} = -(\psi_{i,1} + \psi_{i,1}^\dagger)(\psi_{i,1}^\dagger - \psi_{i,1}) = 2\psi_{i,1}^\dagger \psi_{i,1} - 1 \label{eq:Se1}\\
	S_i^{e_2} &= ic_ib_i^{e_2} = i(\psi_{i,1} + \psi_{i,1}^\dagger)(\psi_{i,2}^\dagger + \psi_{i,2}) = i\psi_{i,1} \psi_{i,2} + i\psi_{i,1} \psi_{i,2}^\dagger + i\psi_{i,1}^\dagger \psi_{i,2} + i \psi_{i,1}^\dagger \psi_{i,2}^\dagger \label{eq:Se2}\\
	S_i^{e_3} &= ic_ib_i^{e_3} = -(\psi_{i,1} + \psi_{i,1}^\dagger)(\psi_{i,2} - \psi_{i,2}^\dagger) = -\psi_{i,1} \psi_{i,2} + \psi_{i,1} \psi_{i,2}^\dagger - \psi_{i,1}^\dagger \psi_{i,2} + \psi_{i,1}^\dagger \psi_{i,2}^\dagger  \label{eq:Se3}
\end{align}
where by the onsite gauge constraint $D_i = b_i^{e_1} b_i^{e_2} b_i^{e_3} c_i = 1$ becomes
\begin{equation} \label{eq:gauge}
    D_i = -(2n_{i,1} - 1)(2n_{i,2} - 1) = 1
\end{equation}
This fermionization is equivalent to both Kitaev's four-majorana representation as well as slave fermion representation \cite{Burnell2011Phys.Rev.B}. To appreciate this point, recall that the Abrikosov decomposition fermionizes a spin-$\frac{1}{2}$ into spinons according to
\begin{align} 
    \tilde{S}_i^{e_1} &= f_{i,\uparrow}^\dagger f_{i,\uparrow} - f_{i,\downarrow}^\dagger f_{i,\downarrow} \label{eq:Sf1}\\
    \tilde{S}_i^{e_2} &= f_{i,\uparrow}^\dagger f_{i,\downarrow} + f_{i,\downarrow}^\dagger f_{i,\uparrow} \label{eq:Sf2}\\
    \tilde{S}_i^{e_3} &= -i(f_{i,\uparrow}^\dagger f_{i,\downarrow} - f_{i,\downarrow}^\dagger f_{i,\uparrow}) \label{eq:Sf3}
\end{align}
with the onsite constraint
\begin{equation}
    n_{i,\uparrow} + n_{i,\downarrow} = 1 \label{eq:const}
\end{equation}
Then if we identify
\begin{equation}
    \psi_{i,1}(\psi_{i,1}^\dagger) \equiv f_{i,\downarrow} (f_{i,\downarrow}^\dagger),~\psi_{i,2}(\psi_{i,2}^\dagger) \equiv f_{\uparrow}(f_{i,\uparrow}^\dagger),~ n_{i,1} \equiv n_{i,\downarrow},~ n_{i,2} \equiv n_{i,\uparrow}
\end{equation}
we would find that Eq. \ref{eq:Sf1} - \ref{eq:Sf3} are close to, but actually not identical to Eq. \ref{eq:Se1}, Eq. \ref{eq:Se2} and Eq. \ref{eq:Se3}.  Indeed, it is straightforward to find their relation: 
\begin{align}
    S_i^{e_1} &= -\tilde{S}_i^{e_1} + (n_{i,1} + n_{i,2} - 1)\\
    S_i^{e_2} &= -\tilde{S}_i^{e_3} + i (\psi_{i,1} \psi_{i,2} + \psi_{i,1}^\dagger \psi_{i,2}^\dagger)\\
    S_i^{e_3} &= -\tilde{S}_i^{e_2} + (\psi_{i,1}^\dagger \psi_{i,2}^\dagger - \psi_{i,1} \psi_{i,2})
\end{align}
However, after a rotation and the application of constraint in Eq. \ref{eq:const} i.e. single particle per site, we immediately see their equivalence $S_i^{e_j} \simeq \tilde{S}_i^{e_j}$. Furthermore, the slave fermion constraint in Eq. \ref{eq:const} and the gauge constraint in Eq. \ref{eq:gauge}, are also equivalent, which can be made explicit by noticing 
\begin{equation}
    D_i = -(2n_{i,1} - 1)(2n_{i,2} - 1) = -2(n_{i,1} + n_{i,2} -1 )^2 + 1 \equiv 1 ~ \Rightarrow~ n_{i,1} + n_{i,2} - 1 = 0
\end{equation}
\begin{figure}
    \includegraphics[width=0.90\textwidth]{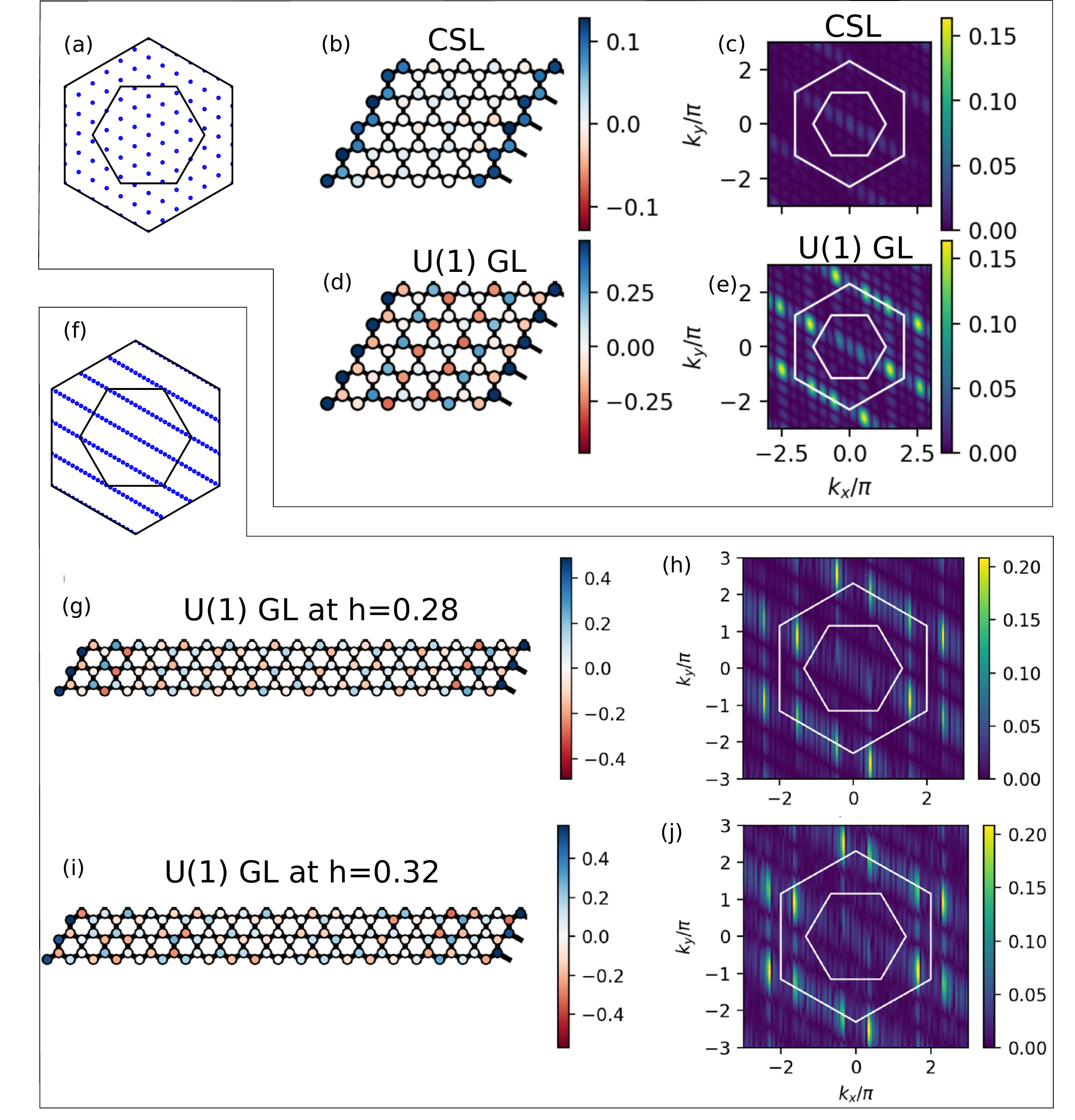}
    \caption{Top block: Results for CSL and gapless U(1) phase obtained by 60-site DMRG. Bottom block: Results for gapless U(1) phase at different fields.
    a) Momentum space resolution of the 60-site cluster. b,c) The magnetization or spinon density modulation of the CSL phase in real and momentum space; and d,e) showing that of the gapless U(1) phase in real and momentum space. f)  Momentum space resolution of the 120-site cluster.
    g,h) Magnetization in the U(1) gapless phase at field $h=0.28$; and i,j) at $h=0.32$, respectively, as well as their momentum space pattern.}
    \label{fig:alt}
\end{figure}
Therefore, the gauge transformed four-majorana representation used in the main text is equivalent to the Abrikosov spinons for spin-$\frac{1}{2}$, which is responsible for the Friedel oscillation in the emergent gapless phase. In this picture, the external magnetic field can be interpreted as a chemical potential for the spinons, which are depleted with increasing field as sketched in \autoref{fig:band}. This mapping is always faithful as long as the on-site constraint is met. In this representation, the Kitaev spin liquid can also be solved in a mean field picture whereby spinons form a weak-pairing p-wave superconductor \cite{Burnell2011Phys.Rev.B}; the gapless intermediate phase is then effectively a neutral metal via a spinon superconductor to spinon metal transition at $h_{c1}$, which is naturally accompanied by a U(1) gauge field and Fermi surface; and the trivial polarized phase is a trivial gapped insulator leading to confinement of spinons.  
Among all three aforementioned phases, only in the U(1) gapless phase does the fermion density (which is equivalent to on-site magnetization amenable to DMRG) oscillate with a length scale $\pi/k_F$ due to the emergence of a gapless Fermi surface \cite{Mross2011Phys.Rev.Ba,Ruan2021Nat.Phys.b}, whereas the CSL and polarized phase do not due to the absence of it.
The real and momentum space modulation of the local $\langle S^{e_1} \rangle$ is shown in \autoref{fig:alt} for different lattice geometry and magnetic fields. 
In the intermediate gapless phase, we note that, the larger lattice with $3 \times 20$ unit cells in the lower panel of \autoref{fig:alt} gives similar pattern to that of the smaller one (top panel and in the main text). For the $h_{c1} < h=0.28 < h_{c2}$ wavefunction, the location of the peak is approximately the same as that measured in the main text, i.e. $q=(-1.5, 1/\sqrt{3})\pi/a$.
For the wavefunction at larger $h_{c1} < h=0.32 < h_{c2}$, the peak is shifted to larger $k$ due to the increasing magnetic field (and thus smaller $k$ when folded into the first Brillouin zone), which is consistent with the spinon picture described in the main text. 

\section{Supplementary Note IV. Analysis of magnetization Bragg peaks in $U(1)$ phase}
\begin{figure}
    \includegraphics[width=0.70\textwidth]{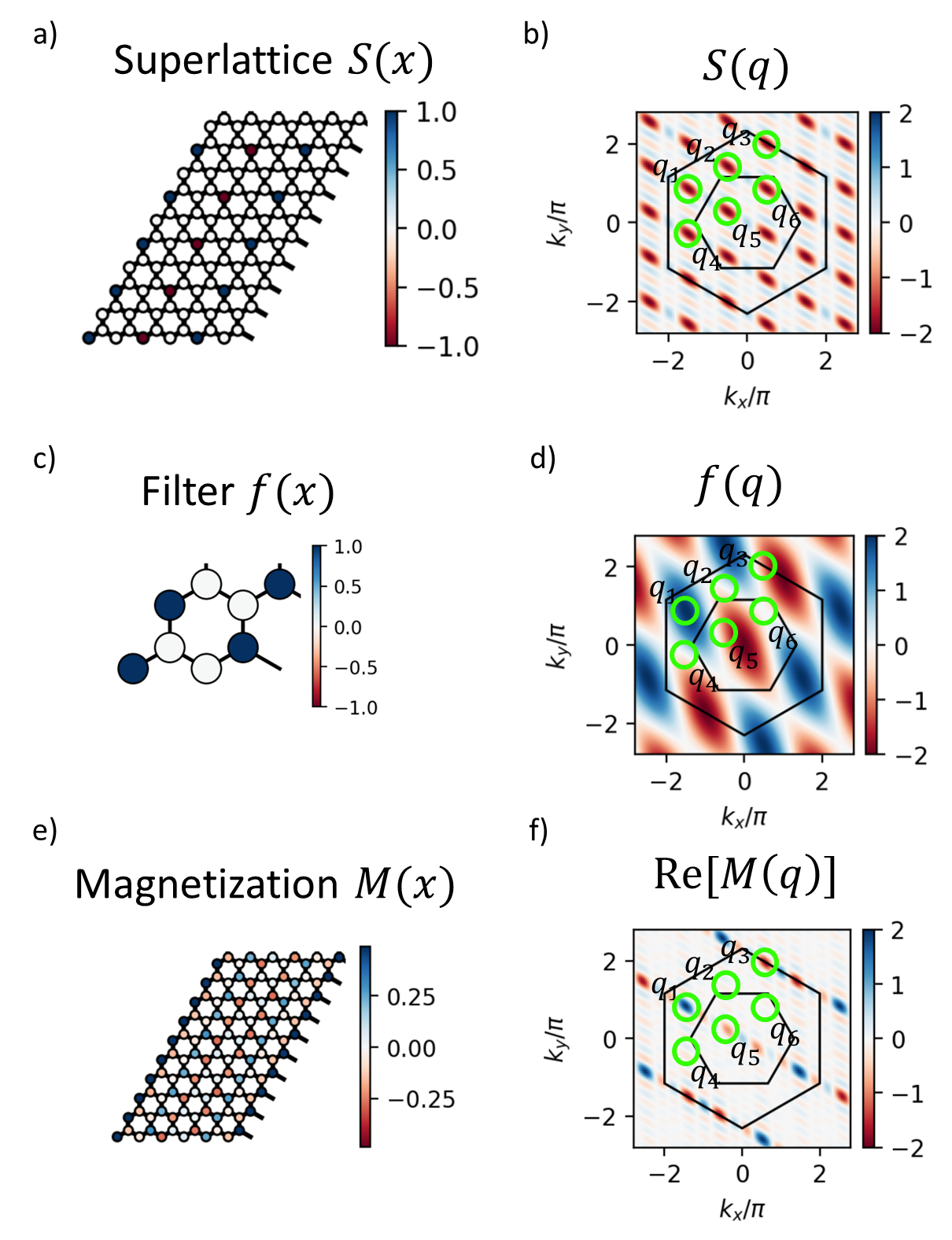}
    \caption{Visualization of Bragg peaks, amplitude predictions, and observed data.
    a) Superlattice $S(x)$ showing antiferromagnetic tiling pattern of filter.
    b) Fourier transform of panel a, with the six Bragg peaks within the extended Brillouin zone marked.
    c) Filter $f(x)$ approximated by discrete 0 and 1 weights.
    d) Fourier transform of panel c.
    e) Numerically obtained magnetization from DMRG lattice, where the lattice has been doubled in the vertical direction as explained in the text.
    f) Real part of Fourier transform of panel e, showing Bragg peaks at the same locations as shown in panel b with varying amplitudes and phases. Imaginary part is negligible and thus not shown.}
    \label{fig:bragg}
\end{figure}

Here we show how the learned filter characterizing the gapless phase predicts the amplitudes of the Bragg peaks in the $\langle S^{e_1}(x) \rangle$ magnetization of the gapless phase.
First, we assume that the magnetization pattern can be formed by a repeated, tiled, filter motif.
Since the applied field is in a perpendicular direction, the tiling pattern must be antiferromagnetic; further, since we performed DMRG with open boundary conditions, the antiferromagnetic tiling must be striped along the open boundary.
The simplest and densest such tiling pattern is shown in \autoref{fig:bragg}(a).
The lattice vectors of the shown superlattice are $\mathbf{d_1} = (2, 0)a$ and $\mathbf{d_2} = (1, \sqrt{3})a$ where $a$ is the lattice constant.
The corresponding reciprocal superlattice vectors are then $\mathbf{b_1} = (1, -\sqrt{3}/3)\pi/a$, $\mathbf{b_2} = (0,2\sqrt{3}/3)\pi/a$, which locate six inequivalent Bragg peaks $q_{1..6}$ within the extended Brillouin zone, as shown in \autoref{fig:bragg}(b).
Then, the expected magnetization pattern $M(\mathbf{x})$ can be interpreted as a convolution of the filter $f(\mathbf{x})$ with the antiferromagnetic superlattice pattern $S(\mathbf{x})$, i.e. $M(\mathbf{x}) = \sum_\mathbf{y} B(\mathbf{x}) S(\mathbf{x}-\mathbf{y})$, where $S(\mathbf{x}) = \sum_{n_1,n_2} (-1)^{n_1}\delta(\mathbf{x} - n_1 \mathbf{d_1} - n_2 \mathbf{d_2})$ is the superlattice pattern.
Then, by the convolution theorem, we have $M(\mathbf{q}) = f(\mathbf{q}) S(\mathbf{q})$, which implies that the Bragg peaks in the magnetization should be located at the reciprocal superlattice vectors, and the peak height should be proportional to $f(\mathbf{q^*})$ where $\mathbf{q^*}$ is the peak location in momentum space.

The size of our DMRG geometry is $6 \times 5$ unit cells, which does not fit an integer number of filter motifs in the vertical direction.
To address this issue, we double the lattice vertically (attaching an identical copy of the same wavefunction) to form a $6 \times 10$ lattice (since we applied periodic boundary conditions in the vertical direction).
The antiferromagnetic superlattice is then $3 \times 5$, which would have Bragg peaks of the same complex phase.
This can be seen by noting that when both dimensions of the superlattice are odd integers, the superlattice has inversion symmetry (i.e. $S(\mathbf{x}) = S(-\mathbf{x})$) and its Fourier transform is thus real.
Further, the geometry is centered at a superlattice site, so the phase contribution to the Fourier transform of all superlattice sites would be the same for all reciprocal lattice wavevectors.

We then compute $f(\mathbf{q})$ by first approximating the filter weights with discrete 0s and 1s, as shown in \autoref{fig:bragg}(c).
Then its Fourier transform is given by the expression
\begin{equation}
    f(\mathbf{q}) = 2\cos(\mathbf{q} \cdot (2\mathbf{a_1}/3-4\mathbf{a_2}/3)) + 2\cos(\mathbf{q} \cdot (2\mathbf{a_1}/3-\mathbf{a_2}/3)),
    \label{eq:ft}
\end{equation}
where $\mathbf{a_1} = (1/2, \sqrt{3}/2)a$, $\mathbf{a_2} = (-1/2, \sqrt{3}/2)a$, and $f(\mathbf{q})$ is plotted in \autoref{fig:bragg}(d).
We also plot the $M(\mathbf{x})$ with doubled lattice and $M(\mathbf{q})$ in \autoref{fig:bragg}(e, f), where Bragg peaks consistent with the reciprocal superlattice shown in \autoref{fig:bragg}(d) can be seen.
For consistency, $M$ has been scaled by an arbitrary factor.
Finally, we compare the theoretical values of $f(\mathbf{q^*})$ and the numerically obtained complex $M(\mathbf{q^*})$ at the six Bragg peaks in \autoref{table:1}.
We find that their ratios and phases follow the same trends, indicating a strong signature of the ML-learned motif in the magnetization.

\begin{table}[h!]
\centering
\caption{Comparison of predicted Bragg peaks $f(\mathbf{q^*})$ following \autoref{eq:ft} with numerically observed complex Bragg peak amplitudes $M(\mathbf{q^*})$, where $M$ has been scaled by an arbitrary factor.}
\begin{tabular}{||c c c||} 
 \hline
 Bragg peak & $f(\mathbf{q^*})$ & $M(\mathbf{q^*})$ \\
 \hline\hline
 $q_1$ & 2 & 2 $\angle$ -0.6$^\circ$\\ 
 $q_2$ & 0 & 0.000  \\
 $q_3$ & -1 & 1.40 $\angle 179.4^\circ$ \\
 $q_4$ & 0 & 0.000 \\
 $q_5$ & -1 & 0.60 $\angle$ -179.3$^\circ$ \\
 $q_6$ & 0 & 0.000 \\
 \hline
\end{tabular}
\label{table:1}
\end{table}

\widetext
\renewcommand\bibsection{\subsection{Supplementary References}}
\bibliography{biblio}